\documentclass[10pt]{article}

\usepackage[dvips]{graphicx}
\usepackage{eepic,epic}
\usepackage{bbm}
\newcommand{\MS}{\mathsf{S}}
\newcommand{\bbH}{\mathbbm{H}}
\newcommand{\C}{\mathbbm{C}}
\newcommand{\R}{\mathbbm{R}}
\newcommand{\Z}{\mathbbm{Z}}
\newcommand{\ONE}{\mathbbm{1}}

\newcommand{\bela}[1]{\begin{equation}\label{#1}}
\newcommand{\ela}{\end{equation}}
\newcommand{\bear}[1]{\begin{array}{#1}}
\newcommand{\ear}{\end{array}}

\newcommand{\X}{\mbox{\boldmath $X$}}
\newcommand{\Y}{\mbox{\boldmath $Y$}}

\renewcommand{\Psi}{\mbox{\boldmath $\psi$}}

\renewcommand{\r}{\mbox{\boldmath $r$}}

\newcommand{\M}{\mathsf{M}}

\newcommand{\as}{\\[.6em]}

\newcommand{\AS}{\\[1.2em]}

\newcommand{\dis}{\displaystyle}
\renewcommand{\i}{\mbox{\rm i}}

\newcommand{\tr}{\,\mbox{tr}\,}

\newcommand{\text}{\textstyle}
\newcommand{\del}{\partial}
\newcommand{\fie}{\varphi}

\begin{document}
\begin{center}
  \Large\bf
  Conformal geometry of the (discrete) Schwarzian Davey-Stewartson 
  II hierarchy\\[8mm]
 \large\sc B.G.\ Konopelchenko\\[2mm]
  \small\sl Dipartimento di Fisica, Universit\`a di Lecce and Sezione INFN, 
  73100 Lecce, Italy\\[3mm]
  \large\sc W.K.\ Schief\\[2mm] 
  \small\sl School of Mathematics, The University of New South Wales,\\
  Sydney, NSW 2052, Australia\\[7mm]
\end{center} 
 
\begin{abstract}
The conformal geometry of the Schwarzian Davey-Stewartson II
hierarchy and its discrete analogue is investigated. Connections with
discrete and continuous isothermic surfaces and generalised Clifford
configurations are recorded. An interpretation of the Schwarzian
Davey-Stewartson II flows as integrable deformations of conformally immersed 
surfaces is given.
\end{abstract}

\section{Introduction}
Due to the (re-)discovery of a variety of important
connections between the differential 
geometry of surfaces and integrable systems, (classical and modern) 
differential geometry has been widely recognised as an integral part of 
soliton theory (see, e.g., \cite{Ten98}-\cite{RogSch02}). However, the 
fundamental nature of geometry in the context of integrable systems is a
subject of ongoing research and recent investigations have uncovered 
unexpected geometric links. For instance, it has been
established that Hirota's master equation \cite{Hir81}
in its Schwarzian form and the associated scalar Schwarzian
Kadomtsev-Petviashvili (SKP) hierarchy are encapsulated in 
Menelaus' classical theorem of plane 
\mbox{geometry~\cite{KonSch02}-\cite{Sch03}}. 

In the present paper, we embark on a study of the geometry of the 
Schwarzian Davey-Stewartson II hierarchy and its discrete analogue, the
quaternionic discrete SKP (qdSKP) equation. We demonstrate that the qdSKP
equation and various associated continuum limits are canonical objects
of conformal (M\"obius) geometry in $\R^4$. In particular, we establish
important connections with both discrete and continuous isothermic surfaces
and generalised Clifford point-circle configurations. 
We also show that the Schwarzian Davey-Stewartson~II hierarchy explicitly 
defines integrable deformations of conformal immersions in~$\R^4$. 

\section{The multicomponent discrete Schwarzian KP equation}

The multicomponent KP hierarchy houses a variety of important soliton 
equations such as the Davey-Stewartson and $N$-wave equations and their 
associated hierarchies \cite{multiKP}. The Schwarzian KP (SKP) hierarchy 
consisting of the singularity manifold equations for the multicomponent KP 
hierarchy has been shown to admit an elegant compact formulation 
\cite{BogKon98}. Indeed, it has been established that if an $N\times N$ matrix 
\bela{E1.1}
  \Phi(t),\quad t = (t_1,t_2,t_3,\ldots)
\ela
depending on an infinite number of `times' $t_n$ constitutes a solution of the
SKP hierarchy then the six solutions
\bela{E1.2}
  \Phi_i = T_i\Phi,\quad \Phi_{ik} = T_iT_k\Phi,\qquad i,k=1,2,3,\quad i\neq k,
\ela
where the `shift' operators $T_i$ are defined by
\bela{E1.3}
  T_i\Phi(t) = \Phi(t + [a_i]) = \Phi\left(t_1+a_i,t_2+\frac{a_i^2}{2},
                                           t_3+\frac{a_i^3}{3},\ldots\right)
\ela
and $a_i={\rm const}$, obey the algebraic 6-point relation
\bela{E2.1}
  \M(\Phi_1,\Phi_{12},\Phi_2,\Phi_{23},\Phi_3,\Phi_{31}) = -\ONE.
\ela
Here, the multi-ratio $\M$ of six matrices $P^1,\ldots,P^6$ is defined by
\bela{E2.1a}
 \bear{l}
  \M(P^1,P^2,P^3,P^4,P^5,P^6)\as 
          \qquad= (P^1-P^2)(P^2-P^3)^{-1}(P^3-P^4)
                               (P^4-P^5)^{-1}(P^5-P^6)(P^6-P^1)^{-1}.
  \ear
\ela
The entire SKP hierarchy may then be retrieved from (\ref{E2.1}) by considering
a canonical limit in which $a_i\rightarrow0$ 
\cite{BogKon98}-\cite{BogKon00}.

As indicated above, the multi-ratio relation (\ref{E2.1})
represents an algebraic superposition formula for six solutions of the
multicomponent SKP hierarchy. As in the scalar case \cite{KonSch02,BogKon99}, 
it also constitutes an algebraic relation (`permutability theorem') for six 
solutions
generated by a variant of the classical Darboux transformation 
\cite{Dar82,MatSal91}. Iterative application of the Darboux transformation
then produces lattices of solutions of the multicomponent SKP hierarchy. In
particular, the multi-ratio relation (\ref{E2.1}) may be interpreted as an 
equation defined on a $\Z^3$ lattice. Thus, in the following, we
regard (\ref{E2.1}) as a discrete equation for a matrix-valued function
\bela{E2.1b}
  \Phi : \Z^3\rightarrow\C^{N,N},\quad (n_1,n_2,n_3)\mapsto\Phi(n_1,n_2,n_3),
\ela
where the indices on $\Phi$ denote translations on the lattice, that is, 
for instance,
\bela{E2.1c}
  \Phi = \Phi(n_1,n_2,n_3),\quad \Phi_1=\Phi(n_1+1,n_2,n_3),\quad
  \Phi_{23}=\Phi(n_1,n_2+1,n_3+1).
\ela

The discrete equation (\ref{E2.1}) has come to be known as the multicomponent 
discrete SKP (dSKP) equation since it encodes the complete multicomponent
SKP hierarchy. Indeed, the genesis of the multicomponent dSKP equation 
implies that any member of 
the multicomponent SKP hierarchy may be obtained from~(\ref{E2.1}) by 
applying an appropriate continuum limit. The integrable nature of the 
multicomponent dSKP equation is inherited from both the original derivation 
(\ref{E1.1})-(\ref{E2.1}) and the construction via Darboux transformations. 
The following analysis is concerned with the conformal geometry of the
quaternionic dSKP equation and its various continuum limits.

\section{The quaternionic dSKP equation}

An important property of the multicomponent dSKP equation is its invariance
under fractional linear transformations of the form
\bela{E2.7}
  \Phi \rightarrow \Phi' = (A\Phi + B)(C\Phi + D)^{-1},
\ela
where $A,B,C$ and $D$ are arbitrary constant matrices. It is therefore
natural to investigate whether contact may be made with conformal
(differential) geometry. Thus, in the remainder of the paper, we assume that
the matrix $\Phi$ takes values in the space of quaternions $\bbH$. The latter
is identified with a four-dimensional Euclidean space $\R^4$ via
\bela{E2.7a}
  \R^4\ni (a,b,c,d)\quad \leftrightarrow\quad
               (a\mathbbm{1} + b\,\mathbbm{i} + c\,\mathbbm{j}
               + d\,\mathbbm{k})\in\mathbbm{H},
\ela
where the matrices $\mathbbm{1},\mathbbm{i},\mathbbm{j},\mathbbm{k}$ are
defined by
\bela{E2.7b}
   \mathbbm{1} = \left(\bear{cc}1&0\\ 0&1\ear\right),\quad
   \mathbbm{i} = \left(\bear{cc}0&-\i\\ -\i&0\ear\right),\quad
   \mathbbm{j} = \left(\bear{cc}0&-1\\ 1&0\ear\right),\quad
   \mathbbm{k} = \left(\bear{cc}-\i&0\\ 0&\i\ear\right).
\ela
In particular, this isomorphism gives rise to the identities
\bela{E3.11c}
  \X^2 = \det X,\quad XX^{\dagger} = (\det X)\ONE,\quad 
  \X\cdot\Y = \frac{1}{2}\tr(XY^{\dagger})
\ela
for any quaternions $X,Y\in\bbH$ and their vectorial analogues $\X,\Y\in\R^4$.
\mbox{Accordingly}, 
in the case $\Phi\in\bbH$, we may refer to (\ref{E2.1}) as the
quaternionic dSKP (qdSKP) equation. The invariance (\ref{E2.7}) 
encodes the group of orientation-preserving conformal transformations in 
$\R^4$ provided that $A,B,C$ and $D$ are quaternions. 

\subsection{The `geometric' continuum limit}

As alluded to in the preceding section, the complete quaternionic SKP hierarchy
may be retrieved from the qdSKP equation via appropriate sophisticated limits.
We embark on a study of the geometric implications of this fact in the next
section. Here, by contrast, we focus on the natural `geometric' limit in which
the differences $\Delta_i\Phi=\Phi_i-\Phi$ are regarded as approximations of
derivatives, that is
\bela{E3.1}
  \Phi_i = \Phi + \epsilon\Phi_{x_i} + O(\epsilon^2),\quad i=1,2,3,
\ela
where $\epsilon$ is a lattice parameter and $\Phi_{x_i}=\del\Phi/\del x_i$.
In the limit $\epsilon\rightarrow0$ and $(x_1,x_2,x_3)=(x,y,z)$, the
qdSKP equation reduces to 
\bela{E3.2}
  \Phi_y\Phi_x^{-1}\Phi_z\Phi_y^{-1}\Phi_x\Phi_z^{-1}=\ONE.
\ela
The latter becomes an identity if the symmetry
\bela{E3.2a}
  \Phi_y + \Phi_z = 0
\ela
is imposed. The scalar analogue of this constraint has been shown to lead to 
conformal maps or (anti-)analytic functions on the complex plane 
\cite{KonSch02}. As in the scalar case, in order to obtain a non-trivial
continuum limit from the qdSKP equation subject to the constraint 
(\ref{E3.2a}), it is required to take into account second-order terms in
the Taylor expansion (\ref{E3.1}) and
consider the terms in the expansion of the
qdSKP equation which are linear in $\epsilon$, that is
\bela{E3.2aa}
 \bear{l}
  \mbox{}\phantom{+}
  \dis\left[\Phi_{xy} + \frac{1}{2}\Phi_{yy}-\Phi_y\Phi_x^{-1}
  \left(\Phi_{xy} + \frac{1}{2}\Phi_{xx}\right)\right]
  \Phi_x^{-1}\Phi_z\Phi_y^{-1}\Phi_x\Phi_z^{-1}\AS
  \mbox{}+
  \dis\Phi_y\Phi_x^{-1}\left[\Phi_{yz} + \frac{1}{2}\Phi_{zz}-\Phi_z\Phi_y^{-1}
  \left(\Phi_{yz} + \frac{1}{2}\Phi_{yy}\right)\right]
  \Phi_y^{-1}\Phi_x\Phi_z^{-1}\AS
  \mbox{}+
  \dis\Phi_y\Phi_x^{-1}\Phi_z\Phi_y^{-1}
  \left[\Phi_{xz} + \frac{1}{2}\Phi_{xx}-\Phi_x\Phi_z^{-1}
  \left(\Phi_{xz} + \frac{1}{2}\Phi_{zz}\right)\right]\Phi_z^{-1} = 0.
 \ear
\ela
Simplification by means of (\ref{E3.2a}) then yields
\bela{E3.4}
  \Phi_{xy} =   \frac{1}{2}\Phi_x\Phi_y^{-1}\Phi_{yy}
              + \frac{1}{2}\Phi_{yy}\Phi_y^{-1}\Phi_x.
\ela
By virtue of the isomorphism $\R^4\cong\bbH$, any solution of the quaternionic
equation~(\ref{E3.4}) gives rise to a surface immersed in $\R^4$ and 
parametrised by the coordinates $x$ and~$y$. Its position vector 
$\r=(r,\mathsf{r})\in\R^4$ is obtained from the decomposition
\bela{E3.6a}
  \Phi = r\ONE + \mathsf{r}\cdot\mathsf{e},
\ela
where the `vector' $\mathsf{e}$ is defined by $\mathsf{e}=(\mathbbm{i},
\mathbbm{j},\mathbbm{k})$. In terms of $\r$, equation (\ref{E3.4}) is 
readily seen to translate into
\bela{E3.10}
  \r_{xy} = \frac{\r_y\cdot\r_{yy}}{\r_y^2}\r_x - 
            \frac{\r_x\cdot\r_{yy}}{\r_y^2}\r_y +
            \frac{\r_x\cdot\r_y}{\r_y^2}\r_{yy}.
\ela
Thus, in the natural geometric continuum limit, the qdSKP equation 
subject to the symmetry constraint (\ref{E3.2a}) governs surfaces 
the position vector of which obeys the second-order
equation (\ref{E3.10}).

By construction, the surfaces defined above belong to conformal 
differential geometry (see, e.g., \cite{Wan92}) due to the invariance of 
(\ref{E3.10}) under the group of conformal (M\"obius) transformations as 
induced by (\ref{E2.7}). These are integrable in
the sense that they are given in terms of solutions of the soliton 
equation~(\ref{E3.4}). In order to proceed, it is now observed that 
(\ref{E3.4}) implies that
\bela{E3.5}
  Q_y = \frac{1}{2}[\Phi_{yy}\Phi_y^{-1},Q],
\ela
where the quaternion $Q$ is defined by
\bela{E3.6}
  Q = \Phi_x\Phi_y^{-1}\Phi_x\Phi_y^{-1}.
\ela
It may therefore be admissible to impose the constraint
\bela{E3.11}
  {(\Phi_x\Phi_y^{-1})}^2 = -\ONE
\ela
since, in this case, relation (\ref{E3.5}) is identically satisfied. In the
generic case, that is $\Phi_x\not\sim\Phi_y$, the constraint (\ref{E3.11}) is
equivalent to the pair
\bela{E3.11a}
  \tr(\Phi_x\Phi_y^{-1}) = 0,\quad \det(\Phi_x\Phi_y^{-1}) = 1
\ela
which, in terms of the position vector $\r$, becomes
\bela{E3.12}
  \r_x^2 = \r_y^2,\quad \r_x\cdot\r_y = 0.
\ela
The latter conditions may be used to cast (\ref{E3.10}) into the form
\bela{E3.11d}
  \r_{xy} = \frac{\r_x\cdot\r_{xy}}{\r_x^2}\r_x + 
            \frac{\r_y\cdot\r_{xy}}{\r_y^2}\r_y
\ela
or, equivalently,
\bela{E3.11e}
  \r_{xy} = a\r_x + b\r_y,
\ela
where the real coefficients $a$ and $b$ are determined by the
constraints (\ref{E3.12}). Thus, it has been established that the surfaces
defined by the system (\ref{E3.12}), (\ref{E3.11e}) constitute a subclass
of the surfaces associated with the continuum limit of the qdSKP
equation considered here.

The geometry of the reduction (\ref{E3.12}), (\ref{E3.11e}) is now readily
revealed. Firstly, the constraints (\ref{E3.12}) are equivalent to demanding
that the first fundamental form of a surface be conformally flat, 
that is
\bela{E3.13}
  d\r^2 = \Omega(dx^2 + dy^2),\quad\Omega=\r_x^2=\r_y^2.
\ela
The coordinates $x$ and $y$ are therefore conformal coordinates \cite{Eis60}.
Secondly, the hyperbolic equation (\ref{E3.11e}) expresses the fact that the
vector $\r_{xy}$ is tangential to the surface so that the coordinates
$x$ and $y$ are conjugate \cite{Eis60}. A surface which may be 
parametrised 
simultaneously in terms of conformal and conjugate coordinates is termed an 
isothermic surface.\footnote{Accordingly, a surface is isothermic if and only 
if its curvature coordinates are conformal modulo a suitable reparametrisation 
$x\rightarrow f(x)$, $y\rightarrow g(y)$ \cite{Eis60}.}
Hence, we conclude that the
continuum limit (\ref{E3.4}) of the qdSKP equation subject to the constraint
(\ref{E3.11}) is associated with isothermic surfaces.
The latter are classical and have been investigated extensively with 
respect to both geometry and integrability (see, e.g., 
\cite{BobSei99,RogSch02} and references therein). The connection with
isothermic surfaces therefore provides a first indication of the fundamental 
nature of the qdSKP equation.

\subsection{The conformal geometry of the qdSKP equation. Discrete isothermic
surfaces}

\begin{figure}
\begin{center}
\setlength{\unitlength}{0.00077489in}
\begin{picture}(5177,3837)(0,-10)
\put(2194,2329){\ellipse{1972}{1972}}
\put(867,1864){\ellipse{1718}{1718}}
\put(1392,2277){\ellipse{1626}{1626}}
\put(3399,1777){\ellipse{3540}{3540}}
\put(1354,2708){\ellipse{1290}{1290}}
\put(1965,3323){\ellipse{984}{984}}
\put(1058,3072){\ellipse{984}{984}}
\put(1932,2802){\ellipse{1416}{1416}}
\put(717,2704){\blacken\ellipse{80}{80}}
\put(1262,2614){\blacken\ellipse{80}{80}}
\put(1682,2149){\blacken\ellipse{80}{80}}
\put(1982,2824){\blacken\ellipse{80}{80}}
\put(2452,3264){\blacken\ellipse{80}{80}}
\put(1557,3069){\blacken\ellipse{80}{80}}
\put(1652,1509){\shade\ellipse{80}{80}}
\put(1477,3334){\shade\ellipse{80}{80}}
\put(1732,1997){$\scriptstyle14$}
\put(2087,2754){$\scriptstyle24$}
\put(2572,3249){$\scriptstyle34$}
\put(517,2704){$\scriptstyle12$}
\put(1302,2982){$\scriptstyle23$}
\put(1357,2589){$\scriptstyle13$}
\put(1602,1359){$\scriptstyle$}
\put(1237,3559){$\scriptstyle1234$}
\put(0522,2007){$\scriptstyle2$}
\put(3182,2007){$\scriptstyle3$}
\put(-60,2007){$\scriptstyle1$}
\put(5182,2007){$\scriptstyle4$}
\put(750,2107){$\scriptstyle124$}
\put(2532,2307){$\scriptstyle134$}
\put(437,3359){$\scriptstyle123$}
\put(2347,3659){$\scriptstyle234$}
\put(1592,1309){$\scriptstyle0$}
\end{picture}
\end{center}
\caption{A $\mathcal{C}_4$ Clifford configuration}
\label{clifford}
\end{figure}
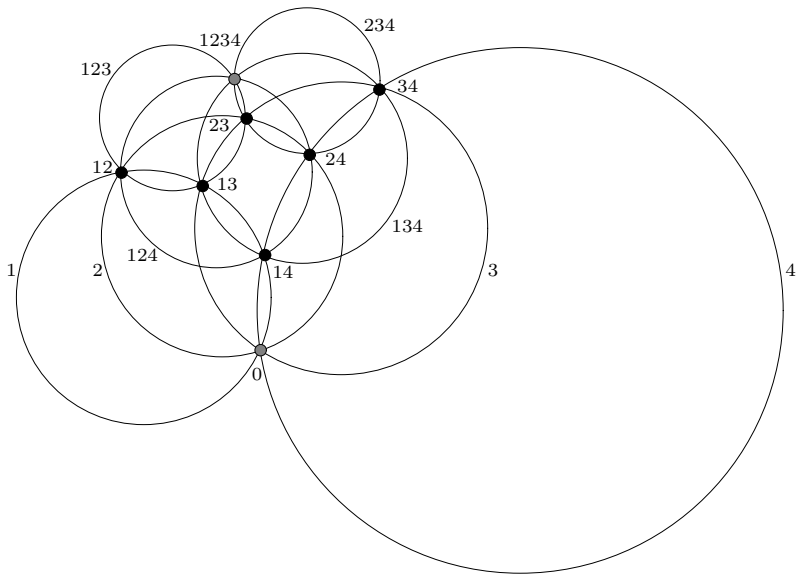

The qdSKP equation has recently been given a geometric interpretation 
\cite{KonSch04} in terms of a novel generalisation of Clifford's classical 
$\mathcal{C}_4$ point-circle configuration. A $\mathcal{C}_4$ Clifford 
configuration is constructed in the following manner \cite{Cli71,Zie41}: 
Consider a
point $P^0$ on a plane and four generic coplanar circles $S^1,S^2,S^3,S^4$ 
passing through $P^0$. The additional six points of intersection are labelled
by $P^{12},P^{13},P^{14},P^{23},P^{24},P^{34}$, where the indices on $P^{ik}$
correspond to those of the intersecting circles $S^i$ and $S^k$ 
(cf.\ Figure \ref{clifford}). 
Any three
circles $S^i,S^k,S^l$ intersect in three points and therefore define a circle
$S^{ikl}$ passing through the points of intersection $P^{ik},P^{il},P^{kl}$.
Clifford's circle theorem then states that, remarkably, the four circles 
$S^{123},S^{124}, S^{134},S^{234}$ meet at a point $P^{1234}$. It is noted in 
passing that
Clifford configurations ($\mathcal{C}_n$) exist for any number of initial
circles $S^1,\ldots,S^n$ passing through a point $P^0$.

In \cite{KonSch02}, it has been shown that the six points $P^{ik}$ of a
$\mathcal{C}_4$ Clifford configuration are algebraically related by the
multi-ratio relation
\bela{E3.13a}
  \M(P^{14},P^{12},P^{24},P^{23},P^{34},P^{13}) = -1
\ela
if the plane is identified with the complex plane so that the points $P^{ik}$
are regarded as complex numbers. The above multi-ratio relation encodes nothing
but Menelaus' classical theorem of plane geometry 
if the point $P^0$ is mapped to 
infinity by a conformal transformation \cite{BraEspGra00}. Here, it is 
important to note
that if we set aside the points $P^0$ and $P^{1234}$ then a $\mathcal{C}_4$
Clifford configuration exhibits the combinatorics of an octahedron if the
six points $P^{ik}$ and eight circles $S^i,S^{ikl}$ are identified with the
vertices and faces of an octahedron respectively. In particular, the 
multi-ratio relation (\ref{E3.13a}) admits the full symmetry group of an 
octahedron acting on the entries $P^{ik}$. Moreover, Ziegenbein \cite{Zie41}
has proven that all circles and points of a $\mathcal{C}_4$
Clifford configuration appear on equal footing in the sense that the angles 
made by four oriented circles passing through a point are the same for all 
eight points. The converse of Ziegenbein's theorem is also valid.

It has been established in \cite{KonSch02} that any `generic' six points 
$P^{ik}$ on the complex plane belong to a $\mathcal{C}_4$ Clifford 
configuration if and only if the multi-ratio relation (\ref{E3.13a}) is 
satisfied. It is therefore natural to inquire as to the geometric significance
of the quaternionic multi-ratio condition
\bela{E3.13b}
  \M(P^{14},P^{12},P^{24},P^{23},P^{34},P^{13}) = -\ONE,
\ela
where the quaternions $P^{ik}$ are regarded as points in a four-dimensional
Euclidean space. It is evident that for any `generic' six points 
$P^{ik}\in\R^4$, the circles $S^i$ and $S^{ikl}$ may still be constructed but,
in general, these do not intersect in any points $P^0$ and $P^{1234}$. 
However, if one imposes the Ziegenbein property then the six points $P^{ik}$ 
can no longer be arbitrary. In fact, it turns out \cite{KonSch04} that the 
Ziegenbein property is equivalent to the quaternionic multi-ratio 
condition.\footnote{Modulo an inversion with respect to a hypersphere.} Thus, 
the latter condition gives rise to generalised $\mathcal{C}_4$ Clifford 
configurations if their definition is based on the Ziegenbein property.

The preceding discussion of Clifford configurations implies that the qdSKP 
equation 
\bela{E3.13c}
  \M(\Phi_1,\Phi_{12},\Phi_2,\Phi_{23},\Phi_3,\Phi_{31}) = -\ONE.
\ela
enshrines `collections' of generalised $\mathcal{C}_4$ Clifford 
configurations. This has been made precise in \cite{KonSch04}. 
On the other hand,
the connection with classical isothermic surfaces as established in 
Section 3.1 raises the question as to 
whether the standard integrable discretisation of isothermic surfaces 
\cite{BobSei99} is related to generalised Clifford configurations. 
A quadrilateral lattice (discrete surface)
\bela{E3.13d}
  \Phi : \Z^2 \rightarrow \R^4\cong\bbH
\ela
is termed isothermic if the quaternionic cross-ratio
\bela{E3.13e}
  Q(\Phi,\Phi_1,\Phi_{12},\Phi_2) = (\Phi-\Phi_1)(\Phi_1-\Phi_{12})^{-1}
                                    (\Phi_{12}-\Phi_2)(\Phi_2-\Phi)^{-1}
\ela
associated with any quadrilateral obeys
\bela{E3.8}
  Q(\Phi,\Phi_1,\Phi_{12},\Phi_2) = -\ONE.
\ela
Since the above quaternionic cross-ratio condition is invariant under 
conformal transformations and any four points in $\R^4$ may be mapped to the
plane by means of an appropriate conformal transformation, the quadrilaterals
are inscribed in circles and their classical (scalar) cross-ratio is
$-1$ \cite{BobSei99}. This is the geometric
content of the quaternionic cross-ratio condition (\ref{E3.8}). 
It is observed that the quaternionic cross-ratio
condition constitutes a natural discretisation of the constraint
(\ref{E3.11}). However, since the quadrilaterals are planar, it also 
discretises the hyperbolic equation (\ref{E3.11e}) so that 
(\ref{E3.8}) may indeed be regarded as a discrete version of the conditions
(\ref{E3.12}), (\ref{E3.11e}) defining classical isothermic 
surfaces~\cite{BobSei99}.

The canonical discrete analogue of the constraint (\ref{E3.2a}) is given by the
translational symmetry
\bela{E3.8a}
  \Phi_{23} = \Phi.
\ela
The latter may be used to eliminate quantities which carry an index 3 from
the qdSKP equation (\ref{E3.13c}). On rearranging terms, one obtains
\bela{E3.8b}
  Q(\Phi_1,\Phi_{12},\Phi_2,\Phi) = T_{\bar{2}}Q(\Phi_{12},\Phi_1,\Phi,\Phi_2),
\ela
where $T_{\bar{2}}f(n_2) = f(n_2-1)$. If the quaternionic
cross-ratio condition (\ref{E3.8}) holds then both cross-ratios in 
(\ref{E3.8b}) are $-\ONE$ due to the symmetry group of the cross-ratio
condition. Accordingly, the pair (\ref{E3.8}), (\ref{E3.8a}) constitutes a 
reduction of the qdSKP equation. In geometric terms, this implies that any
discrete isothermic surface extended to a three-dimensional lattice via the
translational symmetry (\ref{E3.8a}) represents a collection of (degenerate)
generalised $\mathcal{C}_4$ Clifford configurations as introduced in 
\cite{KonSch04} and alluded to in the preceding.
  
\section{Deformation of conformal immersions induced by the Schwarzian 
Davey-Stewartson II hierarchy}

In order to investigate the conformal differential geometry of the continuous 
qSKP hierarchy, we now recall the connection between the qdSKP equation and the
(adjoint) eigenfunctions associated with the quaternionic discrete KP equation,
that is the discrete Davey-Stewartson II equation 
\cite{BogKon98}-\cite{BogKon00}. Thus, it is 
readily verified that the linear system
\bela{F1}
  \phi_2 = \phi_1(\Delta_1\Phi)^{-1}\Delta_2\Phi,\quad
  \phi_3 = \phi_1(\Delta_1\Phi)^{-1}\Delta_3\Phi
\ela
for a quaternionic function $\phi$ is 
compatible modulo the qdSKP equation (\ref{E3.13c}). Hence, one may introduce
a quaternionic function of `separation' $\psi$ according~to
\bela{F2}
  \Delta_i\Phi = \psi\phi_i.
\ela
The compatibility conditions $[\Delta_i,\Delta_k]\Phi=0$ then yield
\bela{F3}
  (\phi_i-\phi_k)\phi_{ik}^{-1} = \psi^{-1}(\psi_k-\psi_i).
\ela
On the one hand, addition of the three relations (\ref{F3}) produces
\bela{F4}
  (\phi_1-\phi_2)\phi_{12}^{-1} + (\phi_2-\phi_3)\phi_{23}^{-1}
  + (\phi_3-\phi_1)\phi_{31}^{-1} = 0
\ela
which constitutes the eigenfunction equation for the quaternionic discrete KP
equation. The latter represents nothing but a discrete 
Davey-Stewartson~II equation so that (\ref{F4}) may be regarded as a discrete
version of the modified Davey-Stewartson~II equation which is known as the
Ishimori equation \cite{Ish84}. 
On the other hand, elimination of the eigenfunction $\phi$ from (\ref{F3})
in a similar manner gives rise to the `adjoint' eigenfunction equation
\bela{F5}
  \psi_1^{-1}(\psi_{31}-\psi_{12}) + \psi_2^{-1}(\psi_{12}-\psi_{23})
  + \psi_3^{-1}(\psi_{23}-\psi_{31}) = 0
\ela
with $\psi$ being an adjoint eigenfunction of the discrete Davey-Stewartson II
equation. It is emphasised that the qdSKP and (adjoint) quaternionic
discrete KP eigenfunction equations are equivalent. For instance, if $\phi$
is an eigenfunction obeying (\ref{F4}) then the linear system (\ref{F3}) for
$\psi$ is compatible and $\psi$ constitutes a solution of the adjoint
eigenfunction equation (\ref{F5}). Moreover, by construction, the existence
of a function $\Phi$ satisfying the defining relations (\ref{F2}) is guaranteed
and $\Phi$ is indeed a solution of the qdSKP equation (\ref{E3.13c}).

The continuum limit to the qSKP hierarchy requires the introduction of the
gauge transformations
\bela{F6}
  \phi \rightarrow A_1^{n_1}A_2^{n_2}A_3^{n_3}\phi,\quad
  \psi \rightarrow \psi A_3^{-n_3}A_2^{-n_2}A_1^{-n_1},
\ela
where the constant matrices $A_i$ constitute non-degenerate diagonal
quaternions. 
In the limit in which the qdSKP equation reduces to the $n$th-order qSKP
equation, the system (\ref{F2}) becomes \cite{BogKon98}-\cite{BogKon00}  
\bela{F7}
  \Phi_x = \psi A_1\phi,\quad \Phi_y = \psi A_2\phi,\quad
  \Phi_t = \mathcal{B}_n(\psi,\phi),
\ela
where $\mathcal{B}_n$ is bilinear in $\psi,\phi$
and their derivatives. If we make the choice $A_1=\ONE$ and $A_2=\mathbbm{k}$ 
and eliminate the eigenfunction $\phi$ from (\ref{F7}) then we obtain the
pair
\bela{F8}
  \Phi_y = N\Phi_x,\quad \Phi_t = \mathcal{L}_n(\Phi),
\ela
where
\bela{F9}
  N = \psi\mathbbm{k}\psi^{-1}
\ela
and $\mathcal{L}_n(\Phi)=\mathcal{B}_n(\psi,\psi^{-1}\Phi_x)$. Accordingly,
$\mathcal{L}_n$ constitutes a linear operator acting on $\Phi$ with 
coefficients
depending on $\psi$ and its derivatives. It turns out that these coefficients 
may be written as differential expressions in $N$ and a scalar auxiliary 
function $\fie$ \cite{BogKon98}-\cite{BogKon00}. The compatibility condition 
for the linear system 
(\ref{F8}) then gives rise to the $n$th-order Ishimori equation. For instance,
in the case $n=2$, the Lax pair (\ref{F8}) assumes the form
\bela{F10}
  \Phi_y = N\Phi_x,\quad \Phi_t = N(\Phi_{xy}-\fie_y\Phi_x-\fie_x\Phi_y)
\ela
and its compatibility condition yields
\bela{F11}
  N_t = NN_{xy} + N_yN_x + \fie_xN_x - \fie_yN_y - (\fie_{xx}+\fie_{yy})N.
\ela
Now, the definition (\ref{F9}) of $N$ implies that
\bela{F12}
  \tr N=0,\quad N^2=-\ONE,\quad N=\MS\cdot\mathsf{e},\quad \MS^2=1
\ela
and hence decomposition of the matrix equation (\ref{F11}) produces 
\bela{F13}
  \bear{c}
    \MS_t = \MS\times\MS_{xy} + \fie_x\MS_x - \fie_y\MS_y\as
    \fie_{xx} + \fie_{yy} + (\MS_x\times\MS_y)\cdot\MS = 0.
  \ear
\ela
The latter represents the Ishimori equation which was first set down in
\cite{Ish84}. It is interesting to note that the `topological charge'
\bela{F14}
  \mathsf{Q} = \frac{1}{4\pi}\int(\MS_x\times\MS_y)\cdot\MS\, dxdy
\ela
is preserved by the Ishimori flow (\ref{F13})$_{1}$. In fact, the topological
charge may be shown to be invariant under all higher-order Ishimori flows. 
It is also remarked that the Ishimori hierarchy is amenable to 
the inverse spectral transform (IST) method \cite{KonMat89,Kon93}.

The geometry of the Schwarzian Davey-Stewartson (SDS) II hierarchy encoded in
the Lax pair (\ref{F8}) for $n=2,3,\ldots$ is unveiled by focussing on the
`scattering problem' (\ref{F8})$_1$ (cf.\ \cite{PedPin98}). Thus, if, for any 
fixed $t$, we identify the 
quaternionic function $\Phi$ with the position vector $\r$ of a surface 
in~$\R^4$ then the properties (\ref{F12})$_{1,2}$ show that the metric of the 
surface is once again given by~(\ref{E3.13})
with $x$ and $y$ being conformal coordinates. Any SDSII flow 
(\ref{F8})$_2$ therefore defines an integrable deformation of conformal
immersions of surfaces in~$\R^4$, where the independent variable $t$ is 
regarded as the deformation parameter. The associated Ishimori flows possess an
infinite set of invariants, the simplest of which is given by the topological
charge (\ref{F14}). Moreover, the invariance of the qdSKP equation under the
M\"obius transformation (\ref{E2.7}) guarantees that the SDSII deformations
are covariant under conformal transformations in $\R^4$. For instance, the
Lax pair (\ref{F10}) is form-invariant under
\bela{F15}
  \Phi\rightarrow \Phi^{-1},\quad N\rightarrow\Phi^{-1}N\Phi,\quad
  \fie\rightarrow\fie -\ln\det\Phi
\ela
corresponding to the composition of an inversion and a reflection in $\R^4$
given by
\bela{F16}
  \r = (r,\mathsf{r})\rightarrow 
  \left(\frac{r}{\r^2},-\frac{\mathsf{r}}{\r^2}\right),\quad
\Omega\rightarrow\frac{\Omega}{\r^4}.
\ela

An important feature of the immersion (\ref{F8})$_1$ and the deformations
(\ref{F8})$_2$ is that these are formulated explicitly in terms of geometric
quantities, namely the position vector $\r$ (or $\Phi$) and the quantity
$N$. The latter is known as the `left normal' of the surface \cite{PedPin98}.
If $\tr\Phi=0$, that is if the surface is embedded in 
$\R^3\cong{\rm Im}\bbH$, then the immersion formula (\ref{F8})$_1$ decomposes
into
\bela{F17}
  \mathsf{r}_y = \MS\times\mathsf{r}_x,\quad \MS\cdot\mathsf{r}_x \ =0
\ela
so that $\MS$ constitutes the unit normal
\bela{F18}
  \MS = \frac{\mathsf{r}_x\times\mathsf{r}_y}{\Omega}
\ela
to the surface. However, in general, $N$ is not normal to the surface and the
constraint $\tr\Phi=0$ is not preserved by the SDSII flows (\ref{F8})$_2$.
In this connection, it is interesting to investigate the stationary points of
the SDSII deformations. For instance, if $\Phi_t=0$ then the flow
(\ref{F10})$_2$ reduces to
\bela{F19}
  \Phi_{xy} = \fie_y\Phi_x + \fie_x\Phi_y
\ela
which is nothing but the hyperbolic equation (\ref{E3.11d}) with
\bela{F20}
  \fie = \frac{1}{2}\ln\Omega.
\ela
Thus, the class of surfaces which is preserved by the deformation associated 
with the Schwarzian Davey-Stewartson II equation coincides with the class of
isothermic surfaces in $\R^4$. The metric of any such surface
is given by
\bela{F21}
  d\r^2 = e^{2\fie}(dx^2 + dy^2)
\ela
which affords an immediate geometric interpretation of the auxiliary 
function~$\fie$. Moreover, the auxiliary equation (\ref{F13})$_2$ constitutes
the Gau{\ss} equation \cite{Eis60}
\bela{F22}
  \fie_{xx} + \fie_{yy} + \mathcal{K}e^{2\fie} = 0,
\ela
where the Gau{\ss}ian curvature $\mathcal{K}$ of the surface is given by
\bela{F23}
  \mathcal{K} = (\MS_x\times\MS_y)\cdot\MS\, e^{-2\fie}.
\ela
The topological charge $\mathsf{Q}$ defined by (\ref{F14}) is therefore
proportional to the total Gau{\ss}ian curvature. Finally, it is noted that
the constraint $\tr\Phi=0$ is compatible with (\ref{F19}) and gives rise to
classical isothermic surfaces in $\R^3$.

\end{document}